\documentstyle[12pt]{article}
\begin{document}
\begin{flushright}
IC/95/118\\
hep-ph/9504274
\end{flushright}
\begin{center}
{\bf THE 331 MODEL WITH RIGHT-HANDED
NEUTRINOS}\\
\vspace{2cm}
{\bf Hoang Ngoc Long}\\
{\it Institute of Theoretical Physics,
National Centre for Natural Science and
Technology,\\
P.O.Box 429, Bo Ho, Hanoi 10000, Vietnam.}\\
\vspace{0.5cm}

and\\
\vspace{0.5cm}

{\it International Centre for Theoretical Physics, Trieste, Italy.}
\vspace{1cm}

Abstract\\
\end{center}
We explore some more consequences of the $SU(3)_L\otimes U(1)_N$
electroweak model with right-handed neutrinos.
By introducing the $Z - Z'$ mixing angle $\phi$, the
{\it exact} physical eigenstates for neutral gauge bosons are
obtained. Because of the mixing, there is a modification to the $Z^1$
coupling
proportional to $\sin\phi$. The data from the $Z$-decay allows us to
fix  the limit for $\phi$ as $-0.0021 \leq \phi \leq 0.000132$.
 From the neutrino neutral current scatterings, we estimate a bound
for the new neutral gauge boson $Z^2$ mass in the range  300 GeV, and
from symmetry-breaking hierarchy a bound for the new charged and neutral
(non-Hermitian) gauge bosons $Y^{\pm}, X^o$ are obtained.

PACS number(s): 12.15.Mm, 12.15.Ff, 12.60.Cn, 12.15.Ji\\

\newpage
\noindent
{\large\bf I. Introduction}\\[0.3cm]
\hspace*{0.5cm}In the Standard Model (SM)~\cite{gaw}, each generation
of fermions is
anomaly-free. This is true for many extensions of the SM as well,
including the popular Grand Unified models~\cite{gg}. In these models,
therefore, the number of generations is completely unrestricted on
theoretical grounds.

Recently, an interesting class of models has been proposed~\cite{svs}
in which each generation is anomalous but different generations are not
exact replicas of one another, and the anomalies cancel when
a number of generations are taken into account, and to be a multiple
of 3. The most economical gauge group which admits  such fermion
representations is $SU(3)_C\otimes
SU(3)_L \otimes U(1)_N$, and it has been proposed by Pisano, Pleitez and
Frampton~\cite{ppf} (for further work on this model, see
Refs.~\cite{fhpp,dng}).
The original model did not have right-handed
neutrinos, but recently we have included them in a non-trivial way in
an interesting modification of the model~\cite{flt}.
We have pointed out that this model is
simpler than the Pisano-Pleitez-Frampton (PPF) model, since fewer Higgs
multiplets are needed in order to allow the fermions to gain masses
and to break the gauge symmetry.

In~\cite{lt} some phenomenological
aspects of the model have been considered. However, these results are based
on an approximate solution for the physical eigenstates of  neutral
gauge bosons.
The purpose of  this paper is to present a further development of the model.
By introducing the $Z-Z'$ mixing angle, the exact physical eigenstates
of neutral gauge bosons are obtained. Based on the data from the
$Z$-decay we fix the mixing angle, similarly, from the neutrino neutral
current
scattering data we estimate  the $Z^2$ boson mass in the range 300 GeV,
which is accessible for direct searches at high energy colliders such
as Tevatron, LHC, NLC, etc.

This paper is organized as follows. In Sec. II we recall some features of the
model
and Yukawa interactions. In Sec. III we
study the gauge boson sector. The charged and neutral currents are given in
Sec. IV. In Sec. V the constraints on the $Z-Z'$ mixing and  masses of the
new gauge bosons  are obtained.
Finally, our conclusions are summarized in the last section.\\[0.3cm]
{\large\bf II. The 331 model and Yukawa interactions}\\[0.3cm]
\hspace*{0.5cm}Like the PPF model, our model
is also based  on the gauge group
\begin{equation}
SU(3)_C\otimes SU(3)_L\otimes U(1)_N.
\label{gg}
\end{equation}
This  model deals with nine leptons and nine quarks. There are three
left- and right-handed neutrinos ($\nu_e, \nu_\mu ,\nu_\tau$),
three charged leptons ($e, \mu, \tau$), four quarks with charge 2/3,
and five quarks with charge  -1/3.

Under the gauge symmetry (1), the three lepton generations
transform as
\begin{equation}
f^{a}_L = \left( \begin{array}{c}
               \nu^a_L\\ e^a_L\\ (\nu^c_R)^a
               \end{array}  \right) \sim (1, 3, -1/3), e^a_R\sim (1,
1, -1),
\label{l}
\end{equation}
where a = 1, 2, 3 is the generation index.

Two of the three quark generations transform identically and one generation
(it does not matter which one)
transforms in a different representation of the gauge group (1):
\begin{equation}
Q_{iL} = \left( \begin{array}{c}
                d_{iL}\\-u_{iL}\\ d'_{iL}\\
                \end{array}  \right) \sim (3, \bar{3}, 0),
\label{q}
\end{equation}
\[ u_{iR}\sim (3, 1, 2/3), d_{iR}\sim (3, 1, -1/3),
d'_{iR}\sim (3, 1, -1/3),\ i=1,2,\]
\[ Q_{3L} = \left( \begin{array}{c}
                 u_{3L}\\ d_{3L}\\ T_{L}
                \end{array}  \right) \sim (3, 3, 1/3),\]
\[ u_{3R}\sim (3, 1, 2/3), d_{3R}\sim (3, 1, -1/3), T_{R}\sim (3, 1, 2/3).\]

It can easily be checked that all gauge anomalies cancel with the
above choice of
gauge quantum numbers. Fermion mass generation and symmetry breaking can be
achieved with just three $SU(3)_{L}$ triplets. We define them by their Yukawa
Lagrangians as follows:
\[
{\cal L}_{Yuk}^{\chi} = \lambda_{1}\bar{Q}_{3L}T_{R}\chi +
 \lambda_{2ij}\bar{Q}_{iL}d^{'}_{jR}\chi^{*} + \mbox{H.c.},
\]
where
\begin{equation}
\chi = \left( \begin{array}{c}
                \chi^o\\ \chi^-\\ \chi^{,o}\\
                \end{array}  \right) \sim (1, 3, -1/3).
\label{h1}
\end{equation}
If $\chi$ gets the vacuum
expectation value (VEV):
\begin{equation}
\langle\chi \rangle^T = (0, 0, \omega/\sqrt{2}),
\label{vevchi}
\end{equation}
then the exotic 2/3 and -1/3 quarks gain
masses and the gauge symmetry is broken to the SM gauge symmetry:
\begin{eqnarray}
&\mbox{SU}(3)_{C}&\hspace*{-0.2cm}\otimes \ \mbox{SU}(3)_{L}\otimes
\mbox{U}(1)_{N}\nonumber \\
&\downarrow      &\hspace*{-0.8cm}\langle \chi \rangle    \\
&\mbox{SU}(3)_{C}&\hspace*{-0.2cm}\otimes \ \mbox{SU}(2)_{L}\otimes
\mbox{U}(1)_{Y},
\nonumber
\label{ssb1}
\end{eqnarray}
where $Y=2N-\sqrt{3}\lambda_8/3$ ($\lambda_8=diag(1, 1,
-2)/\sqrt{3}$). Note that $Y$ is identical
to the standard hypercharge of the SM. Electroweak symmetry breaking
and ordinary fermion mass generation are achieved with two $SU(3)_L$
triplets $\rho , \eta$ which we define through their Yukawa
Lagrangians as follows:
\begin{eqnarray}
{\cal L}_{Yuk}^{\eta} &=& \lambda_{3a}\bar{Q}_{3L}u_{aR}\eta +
 \lambda_{4ia}\bar{Q}_{iL}d_{aR}\eta^{*} + \mbox{H.c.},\nonumber\\
{\cal L}_{Yuk}^{\rho} &=& \lambda_{1a}\bar{Q}_{3L}d_{aR}\rho +
 \lambda_{2ia}\bar{Q}_{iL}u_{aR}\rho^{*} +
G'_{ab}\bar{f}^a_Le^b_R\rho +
 G_{ab}\varepsilon^{ijk}(\bar{f}^a_{L})_i(f^b_L)^c_j
(\rho^{*})_k+ \mbox{H.c.}
\label{yukawa}
\end{eqnarray}
where
\begin{equation}
\rho = \left( \begin{array}{c}
                \rho^+\\ \rho^o\\ \rho^{,+}\\
                \end{array}  \right) \sim (1, 3, 2/3),\
\eta = \left( \begin{array}{c}
                \eta^o\\ \eta^-\\ \eta^{,o}\\
                \end{array}  \right) \sim (1, 3, -1/3).
\label{h2}
\end{equation}
We require the vacuum structure of $\rho , \eta$
\begin{equation}
\langle\rho \rangle^T = (0, u/\sqrt{2}, 0),\
\langle\eta \rangle^T = (v/\sqrt{2}, 0, 0).
\label{vev}
\end{equation}
The last term in Eq.~(\ref{yukawa}) gives the $3 \times 3$ antisymmetric
mass matrix, which has eigenvalues $0, -M, M$. Hence, one of the neutrinos
does not gain mass and the other two are degenerate, at least at the
tree level~\cite{fhpp}. It is easy to see that this term gives interactions
which directly contribute to  lepton-number violation processes such as
neutrinoless double beta decay $(\beta\beta_{0\nu})$ and  neutrino
oscillations.
The VEV $\langle\rho \rangle$ will generate masses for the three
charged leptons, two up-type, one down-type quarks  and two
of the neutrinos will  gain degenerate Dirac masses with one necessarily
massless,
while VEV $\langle\eta \rangle$ will generate masses for the remaining
quarks. The VEVs $\langle\rho \rangle$ and $\langle\eta \rangle$ also
give the electroweak gauge boson masses and results in the symmetry breaking:
\begin{eqnarray}
&\mbox{SU}(3)_{C}&\hspace*{-0.2cm}\otimes \ \mbox{SU}(3)_{L}\otimes
\mbox{U}(1)_{N}\nonumber \\
&\downarrow      &\hspace*{-0.8cm}\langle \chi \rangle   \nonumber \\
&\mbox{SU}(3)_{C}&\hspace*{-0.2cm}\otimes \ \mbox{SU}(2)_{L}\otimes
\mbox{U}(1)_{Y}\nonumber \\
&\downarrow      &\hspace*{-0.8cm}\langle \rho \rangle, \langle
\eta \rangle   \\
&\mbox{SU}(3)_{C}&\hspace*{-0.2cm}\otimes \ \mbox{U}(1)_{Q}.
\nonumber
\label{ssb2}
\end{eqnarray}
Here the electric charge is defined:
\begin{equation}
Q=\frac{1}{2}\lambda_3-\frac{1}{2\sqrt{3}}\lambda_8+N.
\label{charge}
\end{equation}
{\large\bf III. Gauge bosons}\\[0.3cm]
\hspace*{0.5cm}The gauge bosons of this theory form an octet $W^a_\mu$
associated with $SU(3)_L$, an octet $G^a_\mu$ (gluons) with $SU(3)_C$
and a singlet $B_\mu$ associated with $U(1)_N$. It is easy to see
that the massless $G^a_\mu$ gauge bosons associated with $SU(3)_C$
group decouple from the neutral gauge boson mass matrix. Since that
reason, we neglect terms which contain the $G^a_\mu$ gauge
bosons in the covariant derivative. The gauge boson mass matrix arises
from the Higgs boson kinetic term:
\begin{equation}
{\cal L}_{Kinetic} = (D_\mu\chi)^{\dagger}(D^\mu\chi) +
(D_\mu\rho)^{\dagger}(D^\mu\rho) + (D_\mu\eta)^{\dagger}(D^\mu\eta).
\label{kinetic}
\end{equation}
The covariant derivatives are
\begin{equation}
D_\mu  = \partial_\mu  + ig\sum^8_{a=1} W^a_\mu .\frac{\lambda_a}{2}
+ig_N\frac{\lambda^9}{2} N B_\mu ,
\label{derivative}
\end{equation}
where $\lambda^a$(a=1,...,8) are the $SU(3)_L$ generators, and
$\lambda^9=\sqrt{2/3}\  diag(1,1,1)$ are defined such that
$Tr(\lambda^a\lambda^b)=2\delta^{ab}$ and $Tr(\lambda^9\lambda^9)=2$,
and $N$ denotes the $N$ charge for three Higgs multiplets.

The non-Hermitian gauge bosons $\sqrt{2}\ W^+_\mu = W^1_\mu-iW^2_\mu ,
\sqrt{2}\ Y^-_\mu = W^6_\mu- iW^7_\mu ,\sqrt{2}\ X^0_\mu =
W^4_\mu- iW^5_\mu $  have the following masses~\cite{lt}:
\begin{equation}
M^2_W=\frac{1}{4}g^2(u^2+v^2), M^2_Y=\frac{1}{4}g^2(v^2+\omega^2),
M^2_X=\frac{1}{4}g^2(u^2+\omega^2).
\label{mnhb}
\end{equation}
We assume $\langle
\chi\rangle\gg \langle\rho\rangle , \langle\eta\rangle$
such that $M_W\ll M_X , M_Y$. This statement is very
important, because the new gauge bosons must be sufficiently heavy to
keep consistency with low energy phenomenology.

As the triplet scalar $\chi$ acquires a VEV, the symmetry $SU(3)_L
\otimes U(1)_N$ breaks down to $SU(2)_L \otimes U(1)_Y$. By matching
the gauge coupling constants at the $SU(3)_L \otimes U(1)_N$ breaking,
the coupling constant of $U(1)_Y, g'$, is given by
\begin{equation}
\frac{1}{g'^2}=\frac{1}{3g^2}+\frac{6}{g_N^2}.
\label{gmat}
\end{equation}
Eq.~(\ref{gmat}) may be satisfied by a $3-3-1$ mixing angle
$\theta_{3-3-1}$~\cite{lng}
\begin{equation}
g'=\sqrt{3}g\cos\theta_{3-3-1}=\frac{1}{\sqrt{6}}g_N\sin\theta_{3-3-1}.
\label{smc}
\end{equation}
As in the SM we put $g'=g\tan\theta_W$, hence we get finally
\begin{equation}
\frac{g_N}{g}=\frac{3\sqrt{2} \sin\theta_W(M_{Z'})}{\sqrt{3-
4\sin_W^2(M_{Z'})}}.
\label{tn}
\end{equation}
 In this model,
$\sin^2\theta_W$ has to be smaller than $\frac{3}{4}$, while in
the minimal version~\cite{ppf} $\sin^2\theta_W < \frac{1}{4}$.

The neutral (Hermitian) gauge bosons have the $3\times 3$ mass matrix
$M^2$
\begin{equation}
{\cal L}_{mass}=\frac{1}{2}V^TM^2V,
\label{lmass}
\end{equation}
where
\begin{equation}
V^T=(W^3, W^8, B),
\end{equation}
and~\cite{lt}
\begin{equation}
M^{2} = \frac{1}{4}g^2\left( \begin{array}{ccc}
u^2+v^2     &-\frac{1}{\sqrt{3}}(u^2-v^2)   &-\frac{2t}{3\sqrt{6}}(2u^2+v^2) \\
-\frac{1}{\sqrt{3}}(u^2-v^2)& \frac{1}{3}(u^2+v^2+4\omega^2)
 & \frac{2t}{9\sqrt{2}}(2u^2-v^2+2\omega^2)\\
-\frac{2t}{3\sqrt{6}}(2u^2+v^2) & \frac{2t}{9\sqrt{2}}(2u^2-v^2+2\omega^2)
 & \frac{2t^2}{27}(4u^2+v^2+\omega^2)  \\
                            \end{array} \right),
\label{hbm}
\end{equation}
with the notation $t=g_N/g$. This mass matrix can be diagonalized to
obtain the eigenstate fields.

We can identify the photon field $A_\mu$ as well as the massive
bosons $Z$ and $Z'$:
\begin{eqnarray}
A_\mu  &=& s_W  W_{\mu}^3 + c_W\left(-\frac{t_W}{\sqrt{3}}\ W^8_{\mu}
+\sqrt{1-\frac{t^2_W}{3}}\  B_{\mu}\right),\nonumber\\
Z_\mu  &=& c_W  W^3_{\mu} + s_W\left(-\frac{t_W}{\sqrt{3}}\ W^8_{\mu}+
\sqrt{1-\frac{t_W^2}{3}}\  B_{\mu}\right), \nonumber \\
Z'_\mu &=& \sqrt{1-\frac{t_W^2}{3}}\  W^8_{\mu}+\frac{t_W}{\sqrt{3}}\ B_{\mu},
\label{apstat}
\end{eqnarray}
where the mass-squared matrix for { $Z, Z'$} is given by
\[{\cal M}^2 =\left( \begin{array}{cc}
M^2_{Z}     &M_{ZZ'}^2 \\
M_{ZZ'}^2 & M_{Z'}^2\\
                            \end{array} \right),\]
with
\begin{eqnarray}
M_{Z}^2   &=&\frac{g^2}{4 c_W^2}(u^2+v^2)=\frac{M_W^2}{c_W^2},\nonumber \\
M_{ZZ'}^2&=&\frac{g^2}{4c_W^2\sqrt{3-4s_W^2}}
\left[u^2-v^2(1-2s_W^2)\right],\\
\label{mzzp}
M_{Z'}^2 &=&\frac{g^2}{4(3-4s_W^2)}\left[4\omega^2+ \frac{u^2}{c_W^2}
+ \frac{v^2(1-2s_W^2)^2}{c_W^2}\right].
\label{masmat}
\end{eqnarray}
Here we use the following notations: $s_W\equiv \sin\theta_W, c_W\equiv
\cos\theta_W$ and $t_W\equiv \tan\theta_W$.
 From Eq.(23) we see that, the limit for $M_{ZZ'}$ is
\begin{equation}
-\frac{(1-2s_W^2)}{\sqrt{3-4s_W^2}}M_{Z}^2\leq M^2_{ZZ'}\leq
\frac{M_{Z}^2}{\sqrt{3-4s_W^2}}.
\end{equation}
Diagonalizing the mass matrix gives the mass eigenstates $Z^1$ and $Z^2$
which can be taken as mixtures,
\begin{eqnarray}
Z^1  &=&Z\cos\phi - Z'\sin\phi,\nonumber\\
Z^2  &=&Z\sin\phi + Z'\cos\phi.
\end{eqnarray}

The mixing angle $\phi$ is given by
\begin{equation}
\tan^2\phi =\frac{M_{Z}^2-M^2_{Z^1}}{M_{Z^2}^2-M_{Z}^2},
\label{tphi}
\end{equation}
where $M_{Z^1}$ and $M_{Z^2}$ are the {\it physical} mass eigenvalues
\begin{eqnarray}
M^2_{Z^1}&=&\frac{1}{2}\left\{M_{Z'}^2+M_{Z}^2-[(M_{Z'}^2-M_{Z}^2)^2-
4(M_{ZZ'}^2)^2]^{1/2}\right\},\\
M^2_{Z^2}&=&\frac{1}{2}\left\{M_{Z'}^2+M_{Z}^2+[(M_{Z'}^2-M_{Z}^2)^2-
4(M_{ZZ'}^2)^2]^{1/2}\right\}.
\end{eqnarray}
 From Eq.(22) we see that $\phi=0$ if $u^2=v^2(1-2s_W^2)$.
Here $W, Z^1$ correspond to the Standard Model charged and neutral
gauge bosons, and there are new gauge bosons $Y^{\pm}, X^o$, and
$Z^2$.
A fit to precision electroweak observables gives a limit on the
mixing angle (see below) of $-0.0021\leq\phi\leq 0.000132$ and from
the symmetry-breaking hierarchy $\omega\gg u,v$, Eq.~(\ref{mnhb}) and
Eq.~(\ref{masmat}) give us
\begin{equation}
M_{Y^+}\simeq M_{X^o}\simeq \frac{\sqrt{3-4s_W^2}}{2}M_{Z^2}
\simeq 0.72 M_{Z^2}.
\label{masx}
\end{equation}
{\large\bf IV. Charged and neutral currents}\\[0.3cm]
The interactions among the gauge bosons and fermions are read off from
\begin{eqnarray}
{\cal L}_F & = & \bar{R}i\gamma^\mu(\partial_\mu+ig_NB_\mu N)R\nonumber \\
           & + & \bar{L}i\gamma^\mu(\partial_\mu+i\frac{g_N}{\sqrt{6}}
B_\mu N + ig\sum^8_{a=1} W^a_\mu . \frac{\lambda_a}{2})L,
\label{current}
\end{eqnarray}
where $R$ represents any right-handed singlet and $L$ any left-handed
triplet or antitriplet.

The interactions among the charged vector fields with leptons are
\begin{eqnarray}
{\cal L}^{CC}_l = &-& \frac{g}{\sqrt{2}}(\bar{\nu}^a_L\gamma^\mu e^a_LW^+_\mu +
\bar{(\nu^c_R)}^a\gamma^\mu e^a_LY^+_\mu \nonumber \\
&+&\bar{\nu}^a_L\gamma^\mu (\nu^c_R)^aX^0_\mu + \mbox{H.c.}).
\label{ccl}
\end{eqnarray}

For the quarks we have
\begin{eqnarray}
{\cal L}^{CC}_q = &-& \frac{g}{\sqrt{2}}[(\bar{u}_{3L}\gamma^\mu d_{3L}+
\bar{u}_{iL}\gamma^\mu d_{iL})W^+_\mu +
(\bar{T}_{L}\gamma^\mu d_{3L}+\bar{u}_{iL}\gamma^\mu d'_{iL})Y^+_\mu
\nonumber \\
                 &+&(\bar{u}_{3L}\gamma^\mu T_{L}-\bar{d'}_{iL}\gamma^\mu
d_{iL})X^0_\mu + \mbox{H.c.}].
\label{ccq}
\end{eqnarray}
We can see that the interactions with the $Y^+$ and $X^0$ bosons
violate the lepton number (see Eq.(\ref{ccl})) and the weak isospin
(see Eq.(\ref{ccq})).

The electromagnetic current for fermions is the usual one
\begin{equation}
Q_fe\bar{f}\gamma^\mu fA_\mu ,
\end{equation}
where $f$ is any fermion with $Q_f=0, -1, 2/3, -1/3$ and the electric
charge $e$ is identified as follows
\begin{equation}
e = g\sin\theta_W.
\label{egtheta}
\end{equation}

The neutral current interactions can be written in the form
\begin{eqnarray}
{\cal L}^{NC}&=&\frac{g}{2c_W}\left\{\bar{f}\gamma^{\mu}
[a_{1L}(f)(1-\gamma_5) + a_{1R}(f)(1+\gamma_5)]f
Z^1_{\mu}\right.\nonumber\\
             &+&\left.\bar{f}\gamma^{\mu}
[a_{2L}(f)(1-\gamma_5) + a_{2R}(f)(1+\gamma_5)]f Z^2_{\mu}\right\}.
\label{nc}
\end{eqnarray}
The couplings of fermions
with $Z^1$ and $Z^2$ bosons are given as follows:
\begin{eqnarray}
a_{1L,R}(f) &=&\cos\phi\ [T^3(f_{L,R})-s_W^2 Q(f)]\nonumber\\
           &+&c_W^2\left[\frac{3N(f_{L,R})}{(3-4s_W^2)^{1/2}}
-\frac{(3-4s_W^2)^{1/2}}{2c^2_W}Y(f_{L,R})\right]\sin\phi,\nonumber\\
a_{2L,R}(f)&=&-c_W^2\left[\frac{3N(f_{L,R})}{(3-4s_W^2)^{1/2}}
-\frac{(3-4s_W^2)^{1/2}}{2c^2_W}Y(f_{L,R})\right]\cos\phi\nonumber\\
           &+&\sin\phi\ [T^3(f_{L,R})-s_W^2 Q(f)],
\label{vaz}
\end{eqnarray}
where $T^3(f)$ and $Q(f)$ are, respectively, the third component of
the weak isospin and the charge of the fermion $f$. Note that for the exotic
quarks, the weak isospin is equal to zero.
Eqs.~(\ref{vaz}) are valid for both left- and right-handed
currents. Since the value of  $N$ is different for triplets and antitriplets,
the $Z^2$ coupling to left-handed ordinary quarks is different for  the
third family,  and thus flavor changing.

We can also express the neutral current interactions of Eq.~(\ref{nc})
in terms of the vector and axial-vector couplings as follows:
\begin{eqnarray}
{\cal L}^{NC}&=&\frac{g}{2c_W}\left\{\bar{f}\gamma^{\mu}
[g_{1V}(f)-g_{1A}(f)\gamma_5\right] f Z^1_{\mu}\nonumber\\
             &+& \left.\bar{f}\gamma^{\mu}
[g_{2V}(f)-g_{2A}(f)\gamma_5]f Z^2_{\mu}\right\}.
\label{ncva}
\end{eqnarray}
The values of these couplings are:
\begin{eqnarray}
g_{1V}(f)&=&\cos\phi\ [T^3(f_L)-2 s_W^2 Q(f)]\nonumber\\
      &+&\sin\phi\left[
\frac{c_W^2}{(3-4s_W^2)^{1/2}}(3N(f_L)+t^2_W N(f_R))-\sqrt{3-4s^2_W}
\frac{Y(f_L)}{2}\right],\nonumber\\
g_{1A}(f)&=&\cos\phi\ T^3(f_L)\nonumber\\
      &+&\sin\phi\left[
\frac{c_W^2}{(3-4s_W^2)^{1/2}}(3N(f_L)-t^2_W N(f_R))-\sqrt{3-4s^2_W}
\frac{Y(f_L)}{2}\right],\nonumber\\
g_{2V}(f)&=&\cos\phi\left[\sqrt{3-4s^2_W}\frac{Y(f_L)}{2}-
\frac{c_W^2}{(3-4s_W^2)^{1/2}}(3N(f_L)+t^2_W N(f_R))\right] \nonumber\\
       &+&\sin\phi\ [T^3(f_L)-2 s_W^2 Q(f)],\nonumber\\
g_{2A}(f)&=&\cos\phi\left[\sqrt{3-4s^2_W}\frac{Y(f_L)}{2}-
\frac{c_W^2}{(3-4s_W^2)^{1/2}}(3N(f_L)-t^2_W N(f_R))\right] \nonumber\\
       &+&\sin\phi\  T^3(f_L).\nonumber
\end{eqnarray}
The values of $g_{1V}, g_{1A}$ and $g_{2V}, g_{2A}$ are listed in
Tables 1
and 2, where the third generation is assumed to belong to the triplet.
To get some indication as to why the top quark is so heavy, we have to treat
the third generation differently from the first two  as in Refs~\cite{ppf}
and~\cite{lng}. \vspace*{0.3cm}

{\bf TABLE 1}: The $Z^1 \rightarrow f\bar{f}$ couplings in the 331 model
with right-handed neutrinos.\\
\begin{tabular}{|c|c|c|}  \hline
f &$ g_{1V}(f)$ & $g_{1A}(f)$  \\  \hline
$e, \mu, \tau$  & $(-\frac{1}{2}+2s_W^2)(\cos\phi-\frac{\sin\phi}
{(3-4s^2_W)^{1/2}})$&$-\frac{1}{2}(\cos\phi-\frac{\sin\phi}
{(3-4s^2_W)^{1/2}})$\\  \hline
$\nu_e, \nu_{\mu}, \nu_{\tau}$ &$\frac{1}{2}(\cos\phi+\sin\phi
(3-4s^2_W)^{1/2})$ & $\frac{1}{2}(\cos\phi+\sin\phi
(3-4s^2_W)^{1/2})$ \\ \hline
t &$(\frac{1}{2}-\frac{4s_W^2}{3})\cos\phi+\frac{\sin\phi}
{6(3-4s_W^2)^{1/2}}(3+2s_W^2)$& $\frac{1}{2}\cos\phi+\frac{\sin\phi}
{(3-4s_W^2)^{1/2}}(\frac{1}{2}-s_W^2)$\\ \hline
b & $(-\frac{1}{2}+\frac{2s_W^2}{3})\cos\phi+\frac{(3-4s_W^2)^{1/2}}
{6}\sin\phi$ & $-\frac{1}{2}(\cos\phi-\frac{\sin\phi}
{(3-4s^2_W)^{1/2}})$\\  \hline
u,c&$ (\frac{1}{2}-\frac{4s^2_W}{3})(\cos\phi-\frac{\sin\phi}{(3-
4s^2_W)^{1/2}})$&$\frac{1}{2}(\cos\phi-\frac{\sin\phi}{(3-
4s^2_W)^{1/2}})$\\ \hline
d,s&$(-\frac{1}{2}+\frac{2s^2_W}{3})\cos\phi- (\frac{1}{2}-\frac{s^2_W}{3})
\frac{\sin\phi}{(3-4s^2_W)^{1/2}}$&$-\frac{1}{2}\cos\phi-(\frac{1}{2}-s^2_W)
\frac{\sin\phi}{(3-4s^2_W)^{1/2}}$\\ \hline
T&$-\frac{4}{3}s^2_W\cos\phi-(3-7s^2_W)\frac{\sin\phi}{3(3-
4s^2_W)^{1/2}}$&$ -c^2_W\frac{\sin\phi}{(3-
4s^2_W)^{1/2}}$\\ \hline
$d_i'$&$\frac{2}{3}s^2_W\cos\phi+(3-5s^2_W)\frac{\sin\phi}{3(3-
4s^2_W)^{1/2}}$&$ c^2_W\frac{\sin\phi}{(3-
4s^2_W)^{1/2}}$\\ \hline
\end{tabular}
\vspace*{1cm}

{\bf TABLE 2}: The $Z^2\rightarrow f\bar{f}$ couplings.\\
\begin{tabular}{|c|c|c|}  \hline
f &$ g_{2V}(f)$ & $g_{2A}(f)$  \\  \hline
$e, \mu, \tau$  & $(-\frac{1}{2}+2s_W^2)(\sin\phi + \frac{\cos\phi}
{(3-4s^2_W)^{1/2}})$&$-\frac{1}{2}(\sin\phi + \frac{\cos\phi}
{(3-4s^2_W)^{1/2}})$\\  \hline
$\nu_e, \nu_{\mu}, \nu_{\tau}$ &$\frac{1}{2}(\sin\phi - \cos\phi
(3-4s^2_W)^{1/2})$ & $\frac{1}{2}(\sin\phi - \cos\phi
(3-4s^2_W)^{1/2})$ \\ \hline
t &$-\frac{\cos\phi}{6(3-4s_W^2)^{1/2}}(3+2s_W^2)+(\frac{1}{2}-
\frac{4s^2_W}{3})\sin\phi$&
$-(\frac{1}{2}-s_W^2)\frac{\cos\phi}{(3-4s_W^2)^{1/2}} +
\frac{1}{2}\sin\phi$\\ \hline
b & $(-\frac{1}{2}+\frac{2s_W^2}{3})\sin\phi-\frac{(3-4s_W^2)^{1/2}}
{6}\cos\phi$ & $-\frac{1}{2}(\sin\phi+\frac{\cos\phi}
{(3-4s^2_W)^{1/2}})$\\  \hline
u,c&$ (\frac{1}{2}-\frac{4s^2_W}{3})(\sin\phi+\frac{\cos\phi}{(3-
4s^2_W)^{1/2}})$&$\frac{1}{2}(\sin\phi+\frac{\cos\phi}{(3-
4s^2_W)^{1/2}})$\\ \hline
d,s&$(-\frac{1}{2}+\frac{2s^2_W}{3})\sin\phi+ (\frac{1}{2}-\frac{s^2_W}{3})
\frac{\cos\phi}{(3-4s^2_W)^{1/2}}$&$-\frac{1}{2}\sin\phi+(\frac{1}{2}-s^2_W)
\frac{\cos\phi}{(3-4s^2_W)^{1/2}}$\\ \hline
T&$-\frac{4}{3}s^2_W\sin\phi+(3-7s^2_W)\frac{\cos\phi}{3(3-
4s^2_W)^{1/2}}$&$ c^2_W\frac{\cos\phi}{(3- 4s^2_W)^{1/2}}$\\ \hline
$d_i'$&$\frac{2}{3}s^2_W\sin\phi-(3-5s^2_W)\frac{\cos\phi}{3(3-
4s^2_W)^{1/2}}$&$ -c^2_W\frac{\cos\phi}{(3-
4s^2_W)^{1/2}}$\\ \hline
\end{tabular}

We can realize that in the limit $\phi = 0$ the couplings to $Z^1$ of the
ordinary leptons and quarks are the same as in the SM. Furthermore,
the electric charge  defined  in Eq.~(\ref{egtheta}) agrees
with the SM. Because of this, we can test the new
phenomenology beyond the SM. In this
model, the exotic quarks carry electric charges 2/3 and -1/3,
respectively, similarly to ordinary quarks. Consequently, the exotic
quarks can mix with the ordinary ones. This type of mixing gives the
flavor changing neutral currents (FCNCs).
These FCNCs will be induced due to breakdown of the GIM
mechanism. This type of situation has been discussed previously and
bounds on the mixing strengths can be obtained from the
non-observation of FCNC's in the experiments beyond those
predicted by the SM~\cite{ll}.

In the PPF model, the coupling strength of $Z^2$ to quarks is much
stronger than that of leptons due to the factor $1/\sqrt{1-4s_W^2}$.
Therefore, low-energy experiments such as
neutrino-nucleus scattering and atomic parity violation measurements
would be useful to further constrain the model~\cite{dng}. However,
from Tables it is easy to see
that this  does not happen in our model.

In our model, the interactions with the heavy charged and neutral
(non-Hermitian) vector bosons $Y^+, X^o$ violate the lepton
number and the weak isospin. Because of the mixing, the mass eigenstate
$Z^1$ now picks up flavor-changing couplings proportional to $\sin\phi$.
However, since $Z-Z'$ mixing is constrained to be very small,
evidence of 3-3-1 FCNC's can only be probed indirectly at present via
the $Z^2$ couplings.
\\[0.3cm]
{\large\bf V. Constraints on the $Z-Z'$ mixing angle and
the $Z^2$ mass}\\[0.3cm]
\hspace*{0.5cm}There are many ways to get constraints on the
mixing angle $\phi$ and the $Z^2$ mass. Below we present a
simple one. A constraint on the $Z-Z'$ mixing can be followed from the
$Z$-decay data. Hence we now calculate a $Z$ width in this model.\\[0.3cm]
{\it 1. $Z$ decay modes}\\[0.3cm]
The tree-level expression for the partial width for the $Z\rightarrow
f\bar{f}$ where $f=\nu_e$, ...e, u, d,... is given by~\cite{rpp,pll}:
\begin{equation}
\Gamma^{tree}(Z\rightarrow f\bar{f})=\frac{\rho_1 G_F}{6\sqrt{2}\pi}
M^3_{Z^1} N^f_C[(g_{1V}(f))^2+(g_{1A}(f))^2],
\label{zwth}
\end{equation}
where $N^f_C$ is the color factor. From Eq.~(\ref{zwth}) we get
\begin{equation}
\Gamma^{tree}(Z\rightarrow l\bar{l})=\frac{\rho_1 G_F}{6\sqrt{2}\pi}
M^3_{Z^1} \left(\cos\phi-\frac{\sin\phi}{\sqrt{3-4s^2_W}}\right)^2
\left[\frac{1}{4}+(\frac{1}{2}-2s^2_W)^2\right],
\label{zll}
\end{equation}
here $l=e, \mu, \tau$.

To get results consistent with experiments, the QCD and electroweak
radiative corrections have to be included. The weak radiative
corrections that depend upon the assumptions of the electroweak theory
and on the value of the $M_{top}$ and $M_{Higgs}$ are accounted for
by absorbing them into the coupling, which are then called the
{\it effective} coupling $\bar{g}_{1V}$ and $\bar{g}_{1A}$. Then
Eq.~(\ref{zwth}) becomes~\cite{rpp,pll}:
\begin{equation}
\Gamma(Z\rightarrow f\bar{f})=\frac{\bar{\rho_1} G_F}{6\sqrt{2}\pi} M^3_{Z^1}
N^f_C[(\bar{g_1}_V(f))^2+(\bar{g_1}_A(f))^2](1+\delta_{QED})(1+\delta_{QCD}),
\label{rzwth}
\end{equation}
where  $\bar{\rho_1} = 1 + \Delta \rho_1$, $\delta_{QED}=3\alpha Q^2_f/4\pi$
and $\delta_{QCD}=0$ for leptons and $\delta_{QCD}=(\alpha_s/\pi)+
1.409(\alpha_s/\pi)^2-12.805(\alpha/\pi)^3$ for quarks, $\alpha_s$ being
the strong coupling constant at $\mu=M_{Z}$.
Here~\cite{pll} $\sqrt{\rho_1}=
M_W/M_{Z^1} c_W$, and hence from Eqs.~(\ref{mnhb} , 27) we can get its explicit
expression, and see that $\rho_1$ depends on the VEVs and $s_W$. In the limit
$\langle\chi\rangle \gg \langle \rho\rangle, \langle \eta\rangle$, we have
$\rho_1=1$. We will ignore the effects due to the combination of mixing and
radiative corrections since both are very small, i.e.,
\begin{eqnarray}
\bar{g_1}_V(f)&=&\cos\phi\ \bar{g}_V^{SM}(f)+\sin\phi\ \left[
\frac{\bar{c}_W^2}{(3-4\bar{s}_W^2)^{1/2}}(3N(f_L)+\bar{t}^2_W
N(f_R))-\sqrt{3-4\bar{s}^2_W} \frac{Y(f_L)}{2}\right],\nonumber\\
\bar{g_1}_A(f)&=&\cos\phi\ \bar{g}_A^{SM}(f)+\sin\phi\left[
\frac{\bar{c}_W^2}{(3-4\bar{s}_W^2)^{1/2}}(3N(f_L)-\bar{t}^2_W
N(f_R))-\sqrt{3-4\bar{s}^2_W} \frac{Y(f_L)}{2}\right].\nonumber
\end{eqnarray}
The effective coupling constants depend on the fermion $f$ and on the
renormalization scheme~\cite{rpp,amal,bv}:
\[\bar{g}_V^{SM}(f)=\sqrt{\bar{\rho}_{1f}}(T_{3L}(f)-2 Q(f)
\kappa_f\bar{s}^2_W);\
 \bar{g}_A^{SM}(f)=\sqrt{\bar{\rho}_{1f}} T_{3L}(f).\]
For the case $f=b$, where additional vertex corrections are important,
one must replace~\cite{l3} $\bar{\rho_1}$ by $\bar{\rho_b}
=\bar{\rho_1}(1-\frac{4}{3}\Delta\bar{\rho_1})$
and $\bar{s}_W$ by $\bar{s}_W(1+\frac{2}{3}\Delta\bar{\rho_1})$.
Here $\bar{s}_W^2$ is the effective $\sin^2\theta_W$ ~\cite{amal,bar}:
$\bar{s}^2_W = (1 + \Delta\kappa')s^2_W$, and $s^2_W$ is defined
by~\cite{rpp,bar}:
\[s^2_W c^2_W = \frac{\pi \bar{\alpha}(M_Z)}{\sqrt{2}G_F M^2_Z}.\]

By assuming the masses of all the ordinary fermions except the $t$ quark
to be much lighter than the mass of the $Z$ boson  and the masses of the
exotic fermions to be much heavier than the mass of the $Z$ boson,
the total width of the $Z$ boson is given as
\begin{eqnarray}
\Gamma_{total}&=&\Gamma(Z\rightarrow all)=\frac{\bar{\rho_1} G_F}{6\sqrt{2}\pi}
M^3_{Z^1}
\left\{ \cos^2\phi\  \Delta^{SM}_{total}\right.\nonumber\\
             &+&3\sin2\phi\left[G+\frac{\sqrt{3-4\bar{s}_W^2}}{2} -
\frac{D}{4\sqrt{3-4\bar{s}_W^2}}
 +\delta_{QCD}\left(G-\frac{E}{4\sqrt{3-4\bar{s}_W^2}}\right)\right.\nonumber\\
&+&\left.\left.\frac{\alpha}{12\pi}\left(G-\frac{F}{4\sqrt{3-4\bar{s}_W^2}}
\right)\right] +O(\sin^2\phi)\right\},
\label{ztot}
\end{eqnarray}
where
\begin{eqnarray}
D&=&5-\frac{44}{3}\bar{s}_W^2+\frac{272}{9}\bar{s}_W^4;\
E=3-\frac{20}{3}\bar{s}_W^2+\frac{128}{9}\bar{s}_W^4, \nonumber\\
F&=&33-\frac{332}{3}\bar{s}_W^2+
\frac{1808}{9}\bar{s}_W^4;\
G=\frac{\sqrt{3-4\bar{s}_W^2}}{18}(3-2\bar{s}_W^2)-
\frac{(3-4\bar{s}_W^2)^{3/2}}{36}\nonumber
\end{eqnarray}
and
\[\Delta^{SM}_{total}=\sum_{f=\nu,e,u,d,s,c,b}\{[\bar{g}_V^{SM}(f)]^2+
[\bar{g}_A^{SM}(f)]^2\}(1+\delta_{QED}^f)(1+\delta_{QCD}).\]
We get then the ratio
\begin{eqnarray}
R^{331}&=&\frac{\Gamma(Z\rightarrow l\bar{l})}{\Gamma_{total}}=R^{SM}_l
\left\{1-\frac{2\tan\phi}{\sqrt{3-4\bar{s}_W^2}}
\left[1+\frac{3\sqrt{3-4\bar{s}_W^2}}{ \Delta^{SM}_{total}}
\right.\right.\nonumber\\
&\times&\left(G+
\frac{\sqrt{3-4\bar{s}_W^2}}{2}-\frac{D}{4\sqrt{3-4\bar{s}_W^2}}\right.
+\delta_{QCD}\left(G-\frac{E}{4\sqrt{3-4\bar{s}_W^2}}\right)\nonumber\\
&+&\left.\left.\left.\frac{\alpha}{12\pi}\left(G-\frac{F}{4\sqrt{3-4
\bar{s}_W^2}}\right)\right)\right] +O(\tan^2\phi)\right\}.
\label{zrat}
\end{eqnarray}
where $R^{SM}_l$ denotes the SM result: $R^{SM}_l=0.03362$
for~\cite{rpp,ga}\  $ \alpha^{-1}(M_Z)=128.87$,  $\alpha_s(M_Z)=0.118$,
and~\cite{ll91}  $\bar{s}_W^2(M_Z)=0.2333$.
Taking the experimental result in~\cite{rpp}  $\Gamma=(3.367\pm
0.006)\%$, we obtain the limit for the mixing angle
\begin{equation}
-0.0021 \leq \phi\leq 0.000132.
\label{phi}
\end{equation}

As is known, recent results on left-right asymmetry $A_{LR}$
at SLD~\cite{sld} and  $R_b\equiv \Gamma(Z\rightarrow b\bar{b})/
\Gamma(Z\rightarrow  hadrons)$ measured at LEP~\cite{blon}
indicate a possible disagreement at the 2 to 2.5 $\sigma$ level
with the SM prediction ($R_b^{SM}=0.215$ for $M_t$=175 GeV).
If confirmed, this could
indicate new physics coupled in a different way to the third generation.
Therefore, it is interesting to consider $R_b$ in this model.
After some manipulations we get
\begin{eqnarray}
R^{331}_b&=&\frac{\Gamma(Z\rightarrow b\bar{b})}{\Gamma_{hadrons}}=R^{SM}_b
\left\{1-2\tan\phi\left[\frac{9-12\bar{s}_W^2+
8\bar{s}_W^4}{9\sqrt{3-4\bar{s}_W^2}A_b}\right.\right.\nonumber\\
&+&\frac{3}{\left(B+C_h\frac{\alpha}{12\pi}\right)}
\left(G-\frac{E}{4\sqrt{3-4\bar{s}_W^2}}\right.
+\left.\left.\left.\frac{\alpha}{12\pi}\left(G-\frac{F_h}{4\sqrt{3-
4\bar{s}_W^2}}\right)\right)\right]\right.\nonumber\\
&+&\left. O(\tan^2\phi)\right\},
\label{rb}
\end{eqnarray}
where $R^{SM}_b$ is the SM result~\cite{rpp,bv}: $R_b^{SM}=0.215$ and
\begin{eqnarray}
A_b&=&\frac{3}{2}(1-\frac{4}{3}\bar{s}^2_W+\frac{8}{9}\bar{s}^4_W);\
B=\frac{15}{2}-14\bar{s}_W^2+\frac{44}{3}\bar{s}_W^4, \nonumber\\
C_h&=&\frac{33}{2}-38\bar{s}_W^2+\frac{140}{3}\bar{s}_W^4;\
F_h=15-\frac{116}{3}\bar{s}_W^2+\frac{512}{9}\bar{s}_W^4.\nonumber
\end{eqnarray}
Substituting Eq.~(\ref{phi}) into Eq.~(\ref{rb}) we get
\[0.21495\leq R^{331}_b \leq 0.21564.\]
This result still disagrees with the recent experimental
value $R_b=0.2192\pm 0.0018$ measured at LEP~\cite{blon}.
We hope, however, with the inclusion of new heavy particle loop
effects like exotic quarks, Higgs scalars or of new box diagrams,
this result will be improved and consistent with the experimental
data (for recent works on this direction see~\cite{ind}).\\[0.3cm]
{\it 2. Neutrino-electron scattering}\\[0.3cm]
\hspace*{0.5cm}The motivation for focusing on the neutrino neutral current
scatterings is the following: From the theoretical point of view
these reactions
are basic processes free from the complications of strong interactions
and can be used to determine the parameters of the theories. We emphasize
that in the PPF model, these processes are almost the same as in the SM
(for this purpose only neutrino-nucleus scattering and atomic parity
violation, etc, are suitable). The effective four-fermion interactions
relevant to
$\nu$-fermion neutral current processes, in this model, are presented
as follows~\cite{lbn}:
\begin{equation}
-{\cal L}^{\nu f}_{eff}=\frac{\rho_1 G_F}{\sqrt{2}}\bar{\nu}\gamma_\mu
(1-\gamma_5)\nu [C^f_L\bar{f}\gamma^\mu
(1-\gamma_5)f+C^f_R\bar{f}\gamma^\mu (1+\gamma_5)f],
\end{equation}
where
\begin{eqnarray}
C^f_L&=&2\left[(g_{1V}(\nu) + g_{1A}(\nu))(g_{1V}(f) +
g_{1A}(f))+\frac{M_{Z^1}^2}{M_{Z^2}^2}(g_{2V}(\nu) +
g_{2A}(\nu))(g_{2V}(f)+g_{2A}(f))\right],\nonumber\\
C^f_R&=&2\left[(g_{1V}(\nu) + g_{1A}(\nu))
(g_{1V}(f) -g_{1A}(f))+\frac{M_{Z^1}^2}{M_{Z^2}^2}(g_{2V}(\nu) +
g_{2A}(\nu))(g_{2V}(f)-g_{2A}(f))\right].\nonumber
\end{eqnarray}

Then the total cross-sections for $\nu_{\mu}-e$ and
$\bar{\nu}_{\mu}-e$ elastic scattering processes are given, respectively,
\begin{eqnarray}
\sigma(\nu_{\mu}e)&=&\frac{\rho^2_1 m_e E_{\nu} G^2_F}{6\pi}
\left(1-\frac{M_{Z^1}^2}{M_{Z^2}^2}\right)^2\left[\cos2\phi+
\frac{(1-2s_W^2)}{\sqrt{3-4s^2_W}}\sin2\phi\right]^2\left[3(1-2s^2_W)^2+
4s^4_W\right] \nonumber\\
                  &\equiv&\sigma_{SM}(\nu_{\mu}e)
\left(1-\frac{M_{Z^1}^2}{M_{Z^2}^2}\right)^2\left[\cos2\phi+
\frac{(1-2s_W^2)}{\sqrt{3-4s^2_W}}\sin2\phi\right]^2,\\
\label{nes}
\sigma(\bar{\nu}_{\mu}e)&=&\frac{\rho^2_1 m_e E_{\nu} G^2_F}
{6\pi}\left(1-\frac{M_{Z^1}^2}{M_{Z^2}^2}\right)^2\left[\cos2\phi+
\frac{(1-2s_W^2)}{\sqrt{3-4s^2_W}}\sin2\phi\right]^2\left[(1-
2s^2_W)^2+ 12s^4_W\right]\nonumber\\
                          &\equiv&\sigma_{SM}(\bar{\nu}_{\mu}e)
\left(1-\frac{M_{Z^1}^2}{M_{Z^2}^2}\right)^2\left[\cos2\phi+
\frac{(1-2s_W^2)}{\sqrt{3-4s^2_W}}\sin2\phi\right]^2.
\label{anes}
\end{eqnarray}
In the above equations $m_e$ is the mass of the electron and $E_\nu$ is the
energy of the incident (anti)neutrino. From Eq.(46) and
Eq.~(\ref{anes}) we get the ratio of the cross-sections:
\begin{equation}
R=\frac{\sigma(\nu_{\mu}e)}{\sigma(\bar{\nu}_{\mu}e)}=
\frac{3(1-2s^2_W)^2+4s^4_W}{(1-2s^2_W)^2+12s^4_W}= 1.1425,
\end{equation}
which is the same as in the SM~\cite{ah}. In this model,
$\rho_1$ is a free parameter. However, we can follow Degrassi, Fanchiotti
 and Sirlin to put~\cite{pll,ds} $\rho_1=1+\Delta \rho_t$, where
\[\Delta \rho_t\simeq 0.0031\left(\frac{m_t}{100 GeV}\right)^2.\]
Taking an average value for $\phi=-0.00098$,  $m_t=174$ GeV
{}~\cite{ga}, the running $s_W^2=0.21$ and the experimental
results on $\frac{\sigma(\bar{\nu}_\mu e)}{E_\nu}=(1.17\pm 0.206)
\times 10^{-42} cm^2/GeV$ and
$\frac{\sigma(\nu_{\mu}e)}{E_{\nu}}=(1.8\pm 0.32)\times 10^{-42}
cm^2/GeV$  given
in~\cite{ah}, the allowed range of the new gauge boson masses
are $M_{Z^2} \geq 250$ GeV and 330 GeV , respectively . Thus
Eq.~(\ref{masx})
gives a limit for the masses of the gauge bosons $Y^{\pm}, X^o$:
$ M_X  \geq 180$ GeV and 230 GeV, respectively.

In our model, the free parameters are $\sin^2\theta_W, M_{Z^2}$, and
$\phi$
which are constrained from experiment. $M_{Z^1}$ is related by Eq.~(\ref{tphi})
where $M_Z=
M_W/\cos\theta_W$ is the prediction for the $Z$ mass in the absence of
mixing $\phi=0$. It is interesting to consider the special case $\phi=0$.
We have then $M_{Z^1}=M_Z, \rho_1=1$ and $s_W(M_{Z^1})=s_W(M_Z)$.
 From Eq.(46) and Eq.(47) we get bounds for the new gauge bosons $Z^2$ mass:
$M_{Z^2}\geq 270$ GeV and 350 GeV, respectively.
Thus, the only way to get a rigorous bound of $M_{Z^2}$
is through low energy processes as
considered in this paper. Our bounds could be improved
significantly with more precise data.\\[0.3cm]
{\large\bf VI. Discussion}\\[0.3cm]
\hspace*{0.5cm}In this paper, we presented a further development of the 331
model with right-handed neutrinos.
We have shown that this model has some advantages over the original
331 model. Firstly, in the Higgs sector, we need only three Higgs
triplets for generating fermions and gauge bosons  masses as
well as for breaking the gauge symmetry.
Moreover in the limit $\phi = 0$, all couplings of the
ordinary fermions to $Z^1$ boson are the same as in the SM.
In this model there is no limit for the Weinberg angle $\sin^2\theta_W
< \frac{3}{4}$.

In our model, the lepton number is violated both in the Higgs sector
and in the heavy charged and neutral (non-Hermitian) vector bosons
interactions.
We also have flavor-changing neutral currents in the quark sector
coupled to the new $Z^2$ boson. All the heavy bosons have  masses
depending on  $\langle\chi\rangle$ and this VEV is, in principle, arbitrary.

Processes like neutrinoless double beta decay
and neutrino oscillations ($\nu_a \rightarrow \nu_b ;\  a\neq b$), etc, are
typical ones in this model.

Finally, we emphasize
again that  experimental data from the $Z$-decay and
$\bar{\nu}_{\mu}- e ,\  \nu_{\mu}-e$ scattering processes allows us to
estimate the mixing angle $\phi$ and the new gauge boson masses.
To get stronger limits we have to consider other parameters such as
the left-right cross section  asymmetry, $N_\nu$, etc, and we will publish
these results elsewhere~\cite{hnl}.

To summarize, we have shown that because of the $Z-Z'$ mixing there is
a modification to the $Z^1$ coupling proportional to $\sin\phi$,
and the $Z$-decay gives  $-0.0021\leq \phi \leq 0.000132$. The data from
neutrino neutral current elastic scatterings shows that mass of the
new neutral gauge boson  $M_{Z^2}$ is in the range 300 GeV, and from the
symmetry-breaking hierarchy we get:
$M_{Y^+}\simeq M_{X^o}\simeq 0.72 M_{Z^2} \geq 220$ GeV.
We think that new physics can arise at not too high energies.\\[0.3cm]

{\large\bf Acknowledgement}\\[0.4cm]
\hspace*{0.5cm}It is a pleasure to acknowledge
Dr. Robert Foot  for many stimulating and fruitful discussions.
I would like also to thank Prof. G. Altarelli, Prof. G. Senjanovic, Prof.
A. Yu. Smirnov, T. A. Tran for remarks and discussions, and
Prof. Abdus Salam, the International
Atomic Energy Agency and UNESCO for hospitality at the
International Centre for Theoretical Physics, Trieste, where
this work was completed.
I am  grateful to Prof. Tran N. Truong  for continuous help
and Prof. J. Tran Thanh Van for kind attention and partial support.
I thank Profs. C. P. Singh and X. Y. Pham for a careful reading
of the manuscript.\\[0.5cm]

\end{document}